\documentclass[english,a4paper]{article}
\usepackage[T1]{fontenc}
\usepackage[cp1250]{inputenc}
\usepackage{amssymb}
\usepackage{hyperref}
\usepackage{amsfonts}
\usepackage{graphicx}

\begin{document}

\title{Saturation of a spin 1/2 particle by generalized Local control}

\author{F. Mintert\footnote{Freiburg Insitute for Advanced Studies, Albert-Ludwigs University
of Freiburg, Albertstr. 19, 79104 Freiburg Germany}, M. Lapert, Y. Zhang, S. J. Glaser\footnote{Department of Chemistry, Technische Universit\"at M\"unchen, Lichtenbergstrasse 4, D-85747 Garching, Germany} and D. Sugny\footnote{Laboratoire Interdisciplinaire Carnot de Bourgogne (ICB), UMR 5209 CNRS-Universit\'e de Bourgogne, 9 Av. A. Savary, BP 47 870, F-21078 DIJON Cedex, FRANCE, dominique.sugny@u-bourgogne.fr}}

\maketitle

\begin{abstract}
We show how to apply a generalization of Local control design to
the problem of saturation of a spin 1/2 particle by magnetic
fields in Nuclear Magnetic Resonance. The generalization of local
or Lyapunov control arises from the fact that the derivative of
the Lyapunov function does not depend explicitly on the control
field. The second derivative is used to determine the local
control field. We compare the efficiency of this approach with
respect to the time-optimal solution which has been recently
derived using geometric methods.
\end{abstract}
%\pacs{32.80.Qk,37.10.Vz,78.20.Bh} \maketitle
\section{Introduction}
The control of spin systems by magnetic fields has a great
potential for applications in Nuclear Magnetic Resonance (NMR)
\cite{spin}. An example is given by the preparation of a well-defined initial state such as
the saturation process which consists in vanishing
the magnetization vector of a sample by using an adequate pulse
sequence \cite{bydder,patt}. In the setting of NMR spectroscopy and imaging, the saturation control can be used for solvent suppression or
contrast enhancement. Different approaches have been
proposed up to date to solve this control problem. They extend
from intuitive schemes such as the Inversion Recovery (IR) method
\cite{spin} to more elaborate pulse sequences based on geometric optimal
control theory \cite{lapertglaser} or optimization schemes \cite{grapeino,energy,mono,zhang}.
In a recent letter \cite{lapertglaser}, we showed
that a gain of 60 \% in the duration of
the saturation process can be reached when using the time-optimal
solution with respect to the IR one. In general, however, analytic solutions as derived in \cite{lapertglaser} are not available,
and easily accessible numerical methods are required for the control of complex and realistic systems.
Here, we investigate such a method and demonstrate its efficiency through a comparison with time-optimal analytic solutions \cite{lapertglaser}.
Roughly speaking, the present method is an extension of local \cite{local} or Lyapunov control \cite{bookalessandro},
in terms of a so-called Lyapunov function $V$ over the state
space which is minimum (or maximum) for the target state.
In original Lyapunov control theory, a control field that ensures the monotonic decrease of $V$ is constructed with the help of the first time derivative $\dot V$ of $V$.
Under well established general conditions on the
Hamiltonian (see e.g. \cite{lyapunov} for a recent review),
the system converges asymptotically towards the target state.
This approach has been largely explored in quantum mechanics both from the physical
\cite{local,sugawara,lyapunovschirmerex} and mathematical points of view \cite{lyapunovpure}.

In this paper, we will employ a target functional whose first time derivative does {\em not} depend on the available control fields.
In this case, an inspection of $\dot V$ does not help to identify suitable control fields and standard local control theory cannot be used.
Therefore, one has to resort to the second time derivative $\ddot V$, which allows the design of control pulses that speed up the increase of $V$ \cite{mintert}. On the first sight, one might assume that loosing control of $\dot V$ should be a disadvantage.
However, it turns out that it also yields significant advantages:
since control fields typically induce unitary dynamics, the independence of $\dot V$ on the control field implies the invariance of $V$ under the corresponding unitary group. Consequently, the function $V$ is constant for a continuous manifold of quantum states within which the control can induce dynamics. By not enforcing a given path through state space, but leaving the systems the opportunity to find its optimal path with a direction restricted only by the manifolds of constant $V$ permits to not exert unnecessary control, but to leave the system substantial freedom to find its optimal path. Also, $\ddot V$ allows to predict the dynamics of $V$ for longer intervals than $\dot V$, since it extrapolates for two infinitesimal time steps rather than one.
This reduces the time-local character of Lyapunov control, and the efficiency of this generalized approach will be illustrated by different numerical simulations in this work.

The paper is organized as follows. Section \ref{sec2} deals with the description of the system and of the control problem. The description of the generalized local control approach and of the time-optimal one is done in Sec. \ref{sec3}. Different numerical results are finally presented in Sec. \ref{sec4} for relevant situations in NMR. Conclusion and prospective views are given in Sec. \ref{sec5}.
\section{The model system}\label{sec2}
The state of a spin 1/2 particle in an NMR experiment is most conveniently expressed by its Bloch vector with components $M_x$, $M_y$ and $M_z$.
If the radio-frequency magnetic field is in resonance with the spin,
the Bloch equations are given by:
\begin{equation}
   \begin{array}{rcl}
        \frac{dM_x}{dt} & = &   \omega_{y} M_z - \frac{M_x}{T_2} \\
        \frac{dM_y}{dt} & = &  - \omega_{x} M_z - \frac{M_y}{T_2} \\
        \frac{dM_z}{dt} & = & - \omega_{y} M_x  +  \omega_{x} M_y  - \frac{M_z - M_0}{T_1}
    \end{array}
    \label{blocheq}
\end{equation}
where $\omega_{x}$ and $\omega_y$ are the two components of the control field $\vec{\omega}$,
$T_1$ and $T_2$ are two relaxation rates characterizing the incoherent process due to the interaction with the environment and
$M_0$ is the thermal equilibrium magnetization. Since the maximum amplitude $\omega_{max}$ that the control fields $\vec\omega$ can adopt induces a natural time-scale, it is convenient to introduce the scaled variables
\begin{eqnarray*}
\vec{u}&=&2\pi\vec{\omega}/\omega_{max},\\
\tau&=&(\omega_{max}/2\pi)t,\label{eq:scaled_time}\\
\Gamma &=& 2\pi/(\omega_{max}T_2),\\
\gamma&=&2\pi/(\omega_{max}T_1),\\
\vec{x}&=&\vec{M}/M_0\label{eq:scaled_magnetization}.
\label{eq:bloch_orig}
\end{eqnarray*}
Since, in addition, the equations of motions are invariant under rotations around the $z$-axis, we can restrict the entire dynamics to the ($y,z$)-plane by assuming that $\omega_y=0$.
Doing so, Eqs.~(\ref{blocheq}) reduce to
\begin{equation}\label{eqcont}
   % \left\{
    \begin{array}{rcl}
        \dot{y} & = &  - \Gamma y - u z \\
        \dot{z} & = & \gamma(1-z)+ u y
    \end{array} %\right. .
\end{equation}
in terms of the scaled and scalar control field $u=2\pi\omega_x/\omega_{max}$.
\section{Optimization schemes}\label{sec3}
In the following we will take the thermal equilibrium state with $M_z=M_0$ and $M_x=M_y=0$ as initial state.
In terms of the scaled magnetization, this implies that the initial state is located at the north pole of the scaled Bloch sphere. As mentioned below, the goal of the control is to bring the scaled Bloch vector to the center of the sphere in minimum time with the constraint $|u|\leq 2\pi$ on the control field.
\subsection{Generalized local control theory}
\begin{figure}[htbp]
\begin{center}
\includegraphics[width=0.5\columnwidth,angle=90]{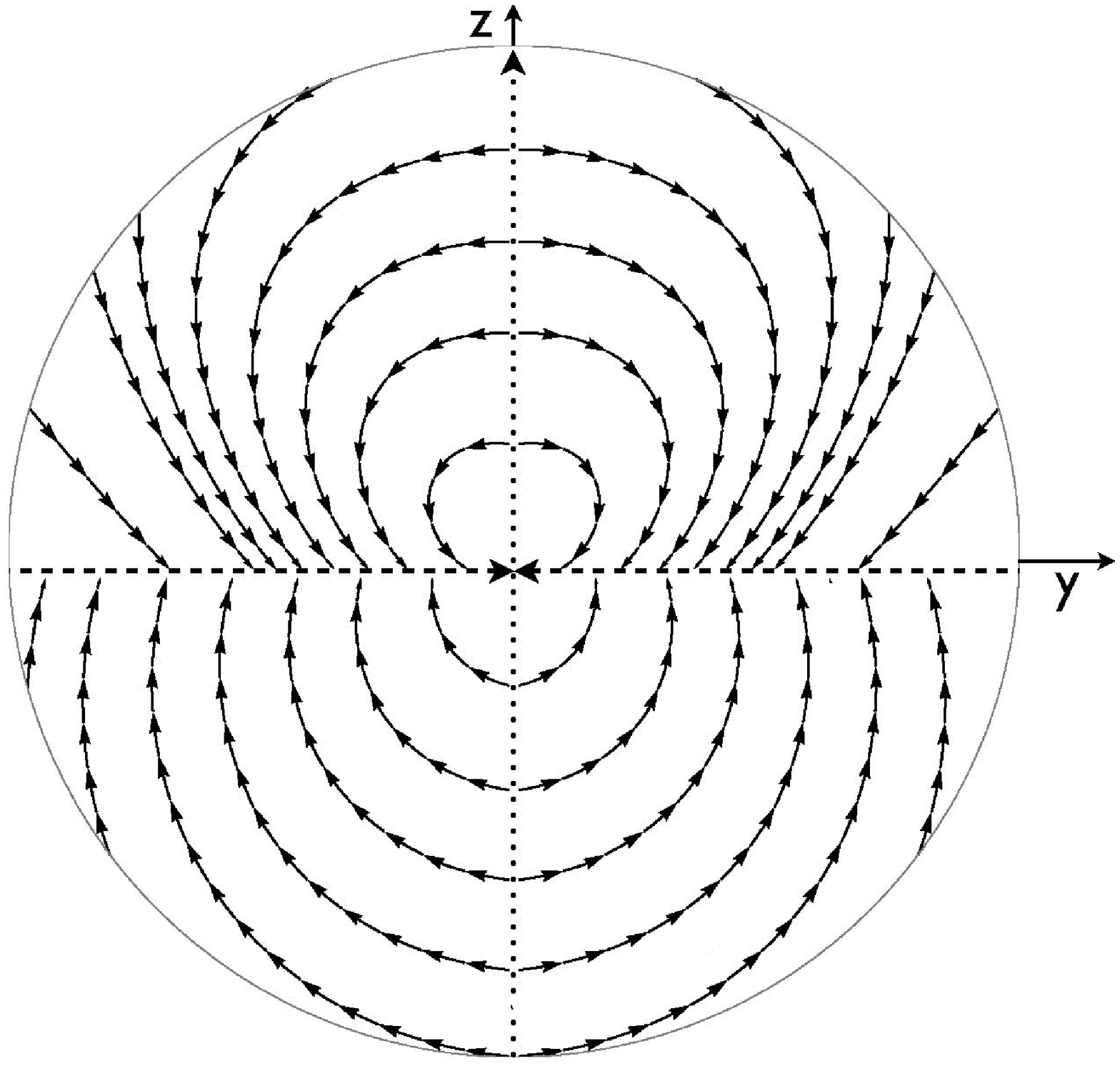}
\caption{Temporal increment $[\dot y,\dot z]$ of the Bloch vector in the $y$-$z$-plane in the presence of local control as given in Eqs.~(\ref{eq:sol1})-(\ref{eq:sol4}).
The dashed arrow depicts the evolution along the line $\mu=0$ according to Eq.~(\ref{eq:sol4}); the dotted arrow corresponds to free evolution according to Eq.~(\ref{eq:sol3}). The curved, solid arrows correspond to Eqs.~(\ref{eq:sol1}) and (\ref{eq:sol2}).}
\label{fig:vectors}
\end{center}
\end{figure}

Since our goal is to reach a vanishing magnetization vector, which is equivalent to a maximum entropy state,
we can choose any entropy-like functional as target. We consider the linear entropy defined as
\begin{equation}
S_l(\varrho)=1-\mbox{Tr}\varrho^2=\frac{1}{2}(1-x^2-y^2-z^2).
\end{equation}
Reaching a maximum entropy in minimum time, i.e. as quickly as possible, can be done by choosing $u$ such that the temporal increment of $S_l$ is maximized.
As anticipated above, the time derivative
\begin{equation}
\frac{dS_l}{dt}=-\Gamma y^2+\gamma(1-z)z,
\end{equation}
of $S_l$ is independent of $u$, which is a direct consequence of the invariance of $S_l$ under coherent dynamics, as generated by $u$.
We, therefore, have to inspect the curvature of $S_l$, which essentially contains the information of how the impact of relaxation changes under the coherent dynamics. In the present case, it reads
\begin{equation}
\frac{d^2S_l}{dt^2}=u\underbrace{y(\gamma-2(\gamma-\Gamma)z)}_{\mu:=\hspace{.5cm}}+\ddot S_l^o,
\end{equation}
where $\ddot S_l^o$ denotes the curvature in the absence of a control field.
Since maximization of the curvature results in maximally fast increase of $\dot S_l$, which in turn implies an acceleration of the increase of entropy,
we will choose, at any instance of time, $u$ such that it maximizes the curvature.
Given that the modulus of $u$ is bounded by $2\pi$, this leads to the choice $u=2\pi$ if $\mu>0$,
and $u=-2\pi$ if $\mu<0$. As long as $\mu$ is finite, this implies that $u$ is constant over a finite period of time so that the equations of motion can be integrated straightforwardly. Care is, however, necessary if $\mu=0$. In this case, one has to distinguish between stable solutions, where the control fields ensure that the condition $\mu=0$ is preserved
and unstable conditions, where the control, or the natural dynamics ensures that the condition $\mu=0$ is satisfied only for an isolated point in time. The latter case is realized for $y=0$ and it corresponds to $u=0$ ({\it i.e.} no control pulse), the system remaining in the line of equation $y=0$ to follow its natural dynamics. The former case is realized for $z=z_0=\gamma/(2(\gamma-\Gamma))$ and requires some more care.
Let us, therefore, step back for a moment from the framework of differential equations and discuss Eqs.~(\ref{eqcont}) in terms of finite time steps.
If we start out with $\mu\leq 0$ than the value of $\mu$ will increase in time since we were assuming the condition $\mu=0$ to be stable.
Due to the finite time steps, however, we might end up in a situation where $\mu\geq 0$.
Now, in turn, $\mu$ will decrease in time, so that we risk to end up in rapid changes of sign of $\mu$,
which results in a rapidly changing sign of the control pulses. Decreasing the size of the time steps will decrease the size of the fluctuations of $\mu$ and increase the frequency with which the control pulse changes sign.
Once this frequency has exceeded all system frequencies sufficiently, we can replace the rapidly oscillating pulse by a time-averaged pulse
\begin{equation}
\bar u(t)=\int d\tilde t\  u(\tilde t) f(t-\tilde t)\ ,
\end{equation}
where $f(t)$ is some function, like a Gaussian, that is positive in some finite domain around $t=0$,
and that decays sufficiently fast outside this domain.
In the limit of infinitesimal time steps, the fluctuation of $\mu$ will vanish, and $\bar u$ is the pulse that has to be applied to preserve the condition $\mu=0$.
The control $\bar u$ needs to be chosen such that
\begin{equation}
\dot{z}=\gamma(1-z)+ \bar u y=0\ .
\end{equation}
This implies that the local pulse solution can be written as
\begin{equation}
\bar u=-\frac{\gamma}{y}(1-z)=-\frac{\gamma}{y}(1-\frac{\gamma}{2(\gamma-\Gamma)})\ .
\label{eq:ubar}
\end{equation}
Note that $\bar{u}\to \pm\infty$ when $y\to 0$ which defines a limit of admissibility due to the bound on the control given by
\begin{equation}
|y|\geq \frac{|\gamma (\gamma-2\Gamma)|}{2\pi (2\Gamma-2\gamma)}.
\end{equation}
This means that for smaller values of $y$, the system cannot follow this horizontal line and a constant control $u=\pm 2\pi$ has to be used.
Together with Eq.~(\ref{eqcont}), the equation of motion for the $y$-component with $u=\bar u$ reads consequently as
\begin{equation}
\dot{y}=  - \Gamma y +\frac{1}{y}\underbrace{\frac{\gamma^2(\gamma-2\Gamma)}{4(\gamma-\Gamma)^2}}_{:=g}\ ,
\end{equation}
which has the solution
\begin{equation}
y_S=\pm \sqrt{e^{-2\Gamma t}(y_0^2-\frac{2g}{\Gamma})+\frac{2g}{\Gamma}}
\end{equation}
where $y_0$ is the initial $y$- coordinate on the horizontal axis. One deduces that $\bar u$ is determined through Eq.~(\ref{eq:ubar}),
and, $\bar u$ in turn, defines $z(t)$ uniquely. One finally arrives at:
%\begin{subequations}
%\begin{equation} u=2\pi\mbox{ for }\mu>0,\label{eq:sol1}\end{equation}%\left\{\begin{array}{ll}\mu>1\mbox{ and }y<0\\\mu<1\mbox{ and %%}y>0\right.\end{array}
%\begin{equation} u=-2\pi\mbox{ for }\mu<0,\label{eq:sol2}\end{equation}%\left\{\begin{array}{ll}\mu<1\mbox{ and }y<0\\\mu>1\mbox{ and %%}y>0\end{array}\right.
%\begin{equation} u=0\mbox{ for }y=0
%%%\mbox{ and }\mu=0
%,\label{eq:sol3}\end{equation}
%\begin{equation} u=\bar u\mbox{ as given by Eqs.~}\eqref{eq:ubar}\mbox{ for }\mu=0\mbox{ and }z=z_0,\label{eq:sol4}\end{equation}
%\end{subequations}
\begin{eqnarray}
& & u=2\pi\mbox{ for }\mu>0,\label{eq:sol1} \\
& & u=-2\pi\mbox{ for }\mu<0,\label{eq:sol2} \\
& & u=0\mbox{ for }y=0\mbox{ and }\mu=0,\label{eq:sol3}\\
& & u=\bar u\mbox{ as given by Eqs.~}(\ref{eq:ubar})\mbox{ for }\mu=0\mbox{ and }z=z_0,\label{eq:sol4}
\end{eqnarray}
and the complete solution is a concatenation of these pieces, as sketched in Fig.~\ref{fig:vectors}. In the following, the control fields of maximum intensity will be called regular or bang, while those such that $|u|<2\pi$ will be said to be singular.

Using this material, we can follow the dynamics starting from the thermal equilibrium ($y=0,z=1$).
This point is an unstable fixed point of the dynamics, {\it i.e.} following the above prescription one should not add any control pulse, the state being stationary. Any small deviation, however, will indicate that a finite control field needs to be applied.
We will, therefore, begin with an infinitesimally displaced initial condition $y(0)=-\varepsilon$, $z(0)=1$.
Then, according to Eq.~(\ref{eq:sol2}) the system is driven with maximum intensity $u=-2\pi$ until the Bloch vector reaches a point with $\mu=0$.
At this point, the control field has to be changed such as to satisfy Eq.~(\ref{eq:sol4}) and the Bloch vector moves along the line $\mu=0$ until its $y$-component vanishes.
Once, this is achieved, the spin will relax to the target state without any control field according to Eq.~(\ref{eq:sol3}). For the moment, we have been assuming that the control field can be chosen strong enough to ensure that the spin remains on the horizontal $\mu=0$- trajectory. Here, it is not the case due to the limitation in the field amplitude, a constant pulse according to Eq.~(\ref{eq:sol2}) has therefore to be applied when $\bar{u}=\pm 2\pi$ along the horizontal line.

\subsection{Optimal control theory}
In this section, we briefly recall the tools of optimal control
theory based on the application of the Pontryagin Maximum
Principle (PMP). The reader is referred to \cite{BS} for a general
application to dissipative quantum systems and to
\cite{lapertglaser,lapert2,zhang} for examples in NMR. The first step of
this method consists in introducing the pseudo-Hamiltonian
$\mathcal{H}$ defined by
\begin{equation}
\mathcal{H}=\vec{P}\cdot (\vec{F}_0(\vec{X})+u\vec{F}_1(\vec{X}))
\end{equation}
where $\vec{X}=(y,z)$ and the vectors fields $\vec{F}_0$ and $\vec{F}_1$ have
respectively the components $(-\Gamma y,\gamma(1-z))$ and
$(-z,y)$. The vector $\vec{P}$ is the adjoint state of coordinates
$(p_y,p_z)$. The PMP states that the optimal trajectories are solutions of the system
\begin{eqnarray*}
& & \dot{\vec{X}}=\frac{\partial \mathcal{H}}{\partial \vec{P}}(\vec{X},\vec{P},v);~\dot{\vec{P}}=-\frac{\partial \mathcal{H}}{\partial \vec{X}}(\vec{X},\vec{P},v)\\
& & \mathcal{H}(\vec{X},\vec{P},v)=\max_{|u|\leq 2\pi}\mathcal{H}(\vec{X},\vec{P},u)\\
& & \mathcal{H}(\vec{X},\vec{P},v)\geq 0.
\end{eqnarray*}
Introducing the switching function $\Phi=\vec{P}\cdot \vec{F}_1$, the PMP leads to two types of situations, the regular and the singular ones. In the regular case, $\Phi(\vec{X},\vec{P})\neq 0$ or vanishes in an isolated point while in the singular case $\Phi$ is zero on a time interval. In the former situation, the corresponding control field is given by $u=2\pi\times\textrm{sign}[\Phi]$. In the latter case where $\phi=\dot{\phi}=\ddot{\phi}=\cdots=0$, we introduce the vector $\vec{V}=(-\gamma+\gamma z-\Gamma z,(-\Gamma+\gamma) y)$ which satisfies $\dot{\phi}=\vec{P}\cdot \vec{V}$. The singular trajectories belong to the set of points where $\vec{V}$ is parallel to $\vec{F}_1$, i.e. to the points such that $\textrm{det}(\vec{F}_1,\vec{V})=0$. A straightforward computation leads to the two lines of equation $y=0$ and $z=z_0$ \cite{lapertglaser}, which shows that the singular set is the same in the local control and optimal approaches. The corresponding optimal singular control is by definition the field such that the dynamics remains on the singular lines. Still this field is identical to local singular control. At this point, it is important to note that the trajectories given by the maximization of $\mathcal{H}$ are only extremal solutions, i.e. candidates to be optimal. Other tools like those introduced in \cite{Bosc2D} have to be used to get optimality results and to select among extremal trajectories the ones which are effectively optimal.

%\iffalse
% In our example, we will see that the local trajectory will be very close but different from the optimal solution. Finally, we point out that such a simple relation between local and optimal control fields is expected in the case of a two-dimensional control problem. It is only for this dimension that both singular set and singular controls can be expressed as a function of the coordinates \cite{Bosc2D}.
%\fi

\section{Comparison of local and optimal control sequences}\label{sec4}

As a specific example to illustrate the efficiency of these two optimization schemes, we consider the control problem of a spin 1/2 particle coupled to a thermal environment as analyzed and experimentally implemented in Ref. \cite{lapertglaser}. In this section, we will consider two examples characterized by different relaxation parameters. In the first case, we have $T_2=8.8$ ms and $T_1=61.9$ ms which leads, with a maximum amplitude $\omega_{max}=2\pi\times 32.3$ Hz, to $\Gamma=3.5$ and $\gamma=0.5$. In the second situation, we choose $T_2=6.2$ ms and $T_1=61.9$ ms which gives with the same maximum amplitude $\Gamma=5$ and $\gamma=0.5$.

\begin{figure}[htbp]
\begin{center}
\includegraphics[width=1\columnwidth]{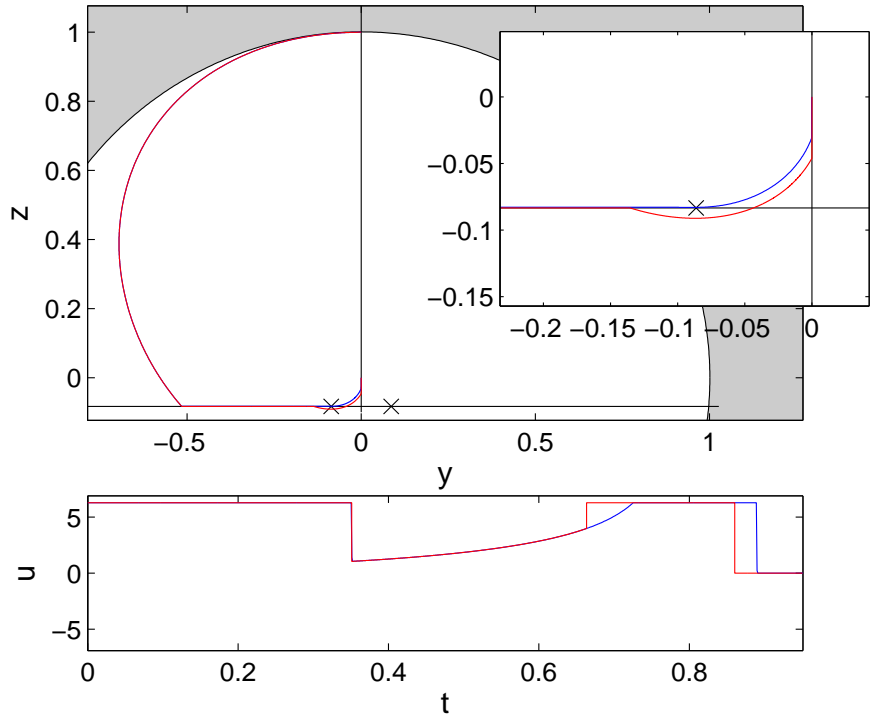}
\end{center}
\caption{(Color Online) (top): Evolution of the magnetization vector in the $(y,z)$-plane for the local control trajectory (blue or black) and the optimal solution (red or gray)
with decoherence-times $T_2=8.8$ ms and $T_1=61.9$.
The insert is a zoom of the trajectories near the singular horizontal line. The cross indicates the position of the limit of admissibility of the control along the horizontal axis. The bottom panel displays the corresponding control fields. \label{fig2}}
\end{figure}

\begin{figure}[htbp]
\begin{center}
\includegraphics[width=1\columnwidth]{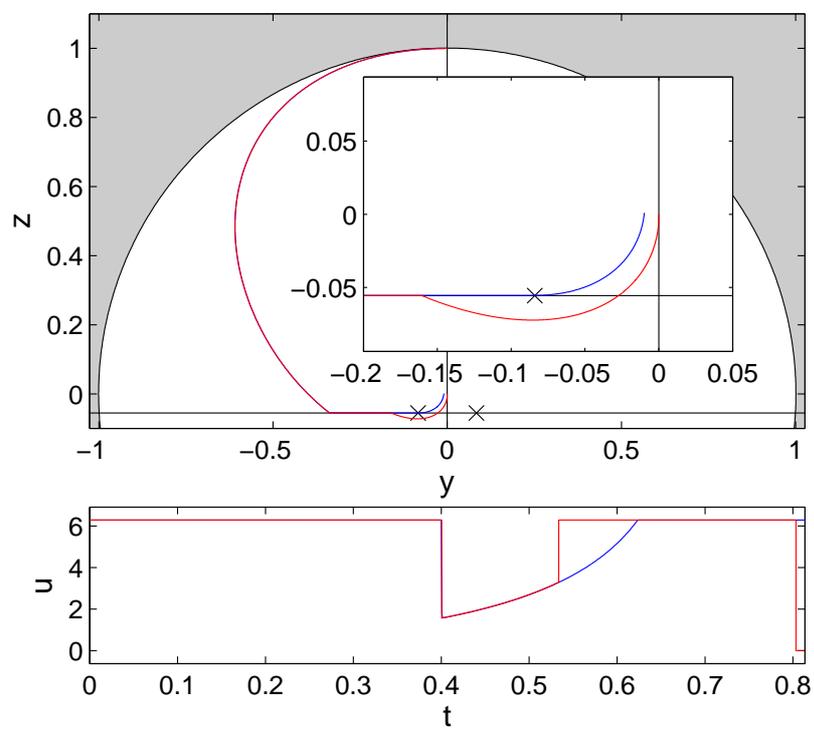}
\end{center}
\caption{(Color Online) Evolution of the magnetization vector similarly to Fig. \ref{fig2} but with decoherence-times $T_2=6.2$ ms and $T_1=61.9$.\label{fig3}}
\end{figure}

Let us compare the time-optimal solution as computed in \cite{lapertglaser,BS} with the solutions obtained with the generalized local control theory. The numerical results are represented in Fig. \ref{fig2} and \ref{fig3} for the first and second cases, respectively. In the two cases, the time optimal solution is composed of two bang pulses of maximum amplitude $2\pi$ and two singular controls:
starting from the initial point of coordinates ($y=0$,$z=1$), the first bang pulse drives the spin up to the horizontal singular line where $\mu$ vanishes. The consecutive singular control ensures that the $z$-component remains constant until  $y=y_c$, when the second bang pulse is applied. The position $y_c$ is computed to ensure that the switching function $\phi$ and its first derivative $\dot{\phi}$ remain continuous along the optimal sequence. The values of $y_c$ are equal to -0.1605 and -0.1356 in the first and second cases. The second bang pulse increases the $z$-component of the magnetization vector until its $y$-component vanishes. The final part of the optimal sequence is a zero control along the vertical singular axis $y=0$ which allows to reach the target state. With the chosen relaxation parameters $\gamma$ and $\Gamma$, the time-optimal sequence has a duration of 0.95050 and 0.97  in dimensionless units [defined above in Eq.~(\ref{eq:scaled_time})] in the first and second cases. When there is no bound on the control, the second bang does not exist and the optimal sequence is formed by a bang arc followed by the two singular trajectories along the horizontal and vertical lines. The solution of the generalized local control is very similar. Exactly like the time-optimal solution, it begins with a bang pulse followed by a singular pulse ensuring that $\mu$ vanishes, but this pulse is maintained longer than in the time-optimal solution. Only at $y=-0.0862$ and $y=-0.0840$ where $|u|=2\pi$, the limited control intensity does not permit to maintain the condition $\mu=0$ and a bang pulse has to be applied. In the unbounded situation, note that the local and optimal solutions are identical. In the bounded case, if the second bang arc intersects the vertical $z$- axis, the dynamics relaxes without any control until its Bloch vector vanishes. In the case of Fig. \ref{fig2}, this leads to an overall control duration of 0.95098 which is slightly larger than the one of the time-optimal solution, the difference is of the order of $4.81\times 10^{-4}$. Whereas local control does not reproduce the time-optimal solution exactly, we can conclude in this case that the longer duration of the former is negligible for all practical purposes. This conclusion is not true for all the values of the relaxation parameters. In particular in Fig. \ref{fig3}, since there is no intersection between the second bang pulse and the $z$- axis, we observe that the local control approach does not manage to reach the target, the minimum distance being equal to $9.9\times 10^{-4}$.

\begin{figure}[htbp]
\begin{center}
\includegraphics[width=1\columnwidth]{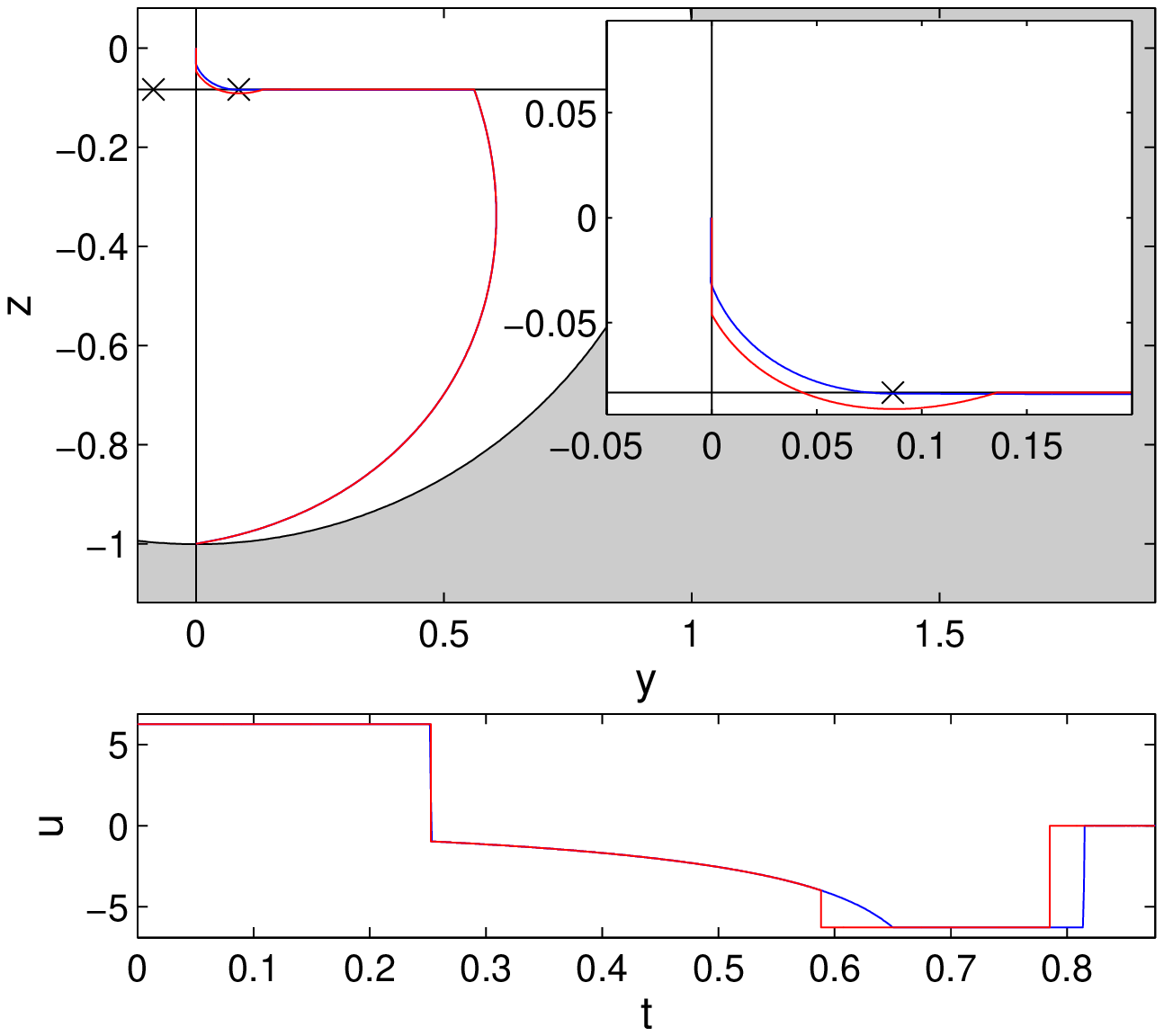}
\end{center}
\caption{(Color Online) Evolution of the magnetization vector with decoherence-times $T_2=6.2$ ms and $T_1=61.9$ as in Fig. \ref{fig2}, but for an initial state corresponding to the south pole of the Bloch sphere.\label{fig4}}
\end{figure}

Figure \ref{fig4} gives a third example where the initial point of the dynamics is the south pole of the Bloch sphere, the relaxation rates being the ones of the first example. If we assume that the initial point is exactly on the $z$- axis then the local solution consists of letting act the relaxation with $u=0$ up to the center of the Bloch ball. Since this axis is unstable, any small deviation from this point will induces a bang control of amplitude $\pm 2\pi$ up to the horizontal singular line. A second bang pulse is applied when the limit of admissibility of the control is reached and a zero control finally used up to the target state. The optimal control sequence is quite similar, except for the second bang which is applied earlier than in the local control strategy. We observe in Fig. \ref{fig4} that the local and the optimal solutions are very close to each other with a control duration respectively of 0.8758 and 0.8753, while if $u=0$ this time is of 1.3861.

\section{Conclusion and perspectives}\label{sec5}

The close-to-optimal performance of the generalized time-local control pulses reflect the potential of this new approach which can efficiently replace the standard local control design when this strategy cannot be used. Whereas we have seen that sufficiently strong control typically permits to reach the target, it can also be missed with weak control fields.
This opens up the general question of convergence in the present generalized framework of Lyapunov control.
In standard Lyapunov control theory there is a well-established theory of convergence, based, e.g., on the concept of the Lasalle
invariant set on which the derivative of the Lyapunov function
vanishes \cite{lasalle}. In addition, necessary and sufficient
conditions on the Hamiltonian of the system have been derived to
ensure the converge towards the target state \cite{lyapunov}.
Given the success of these tools for standard Lyapunov control theory it seems within reach to also develop rigorous concepts to predict convergence properties of the present extension.\\ \\
\noindent \textbf{Aknowledgment} S.J.G. acknowledges support from the DFG
(GI 203/6-1), SFB 631. S. J. G. thank the Fonds der
Chemischen Industrie. F.M. gratefully acknowledges support of the European Research Council (259264).\\ \\
%\section*{References}

\end{document}